\documentclass[sigplan,nonacm]{acmart}

\setcopyright{none}
\copyrightyear{2026}
\acmYear{2026}
\acmDOI{XXXXXXX.XXXXXXX}

\acmConference[ASPLOS 2027]{International Conference on Architectural Support for Programming Languages and Operating Systems}{April 11-- 15, 2027}{Crete, Greece}

\acmISBN{978-X-XXXX-XXXX-X/XX/XX}

\renewcommand\footnotetextcopyrightpermission[1]{}

\usepackage{booktabs}
\usepackage{tikz}
\usepackage{comment}
\usepackage{amsmath}
\usepackage{caption}
\usepackage{subcaption}
\usepackage{graphicx}
\usepackage{float}
\usepackage{algorithm}
\usepackage{algorithmicx}
\usepackage[noend]{algpseudocode}
\algnewcommand{\LineComment}[1]{\State \(//\) #1}
\usepackage[normalem]{ulem}
\usepackage{bbding}
\usepackage{todonotes}
\usepackage{filecontents}
\usepackage{enumitem}
\usepackage[bottom]{footmisc}
\usepackage[]{hyperref}
\usepackage{cleveref}
\usepackage{pifont}
\usepackage{multirow}
\usepackage{xcolor}
\usepackage{xparse}
\usepackage{outlines}
\usepackage{lipsum}
\usepackage{svg}

\definecolor{LightGray}{gray}{0.9}
\definecolor{FadedBanana}{RGB}{255,255,191}
\definecolor{DeepChalk}{RGB}{255,191,191}
\definecolor{FadedFlora}{RGB}{191,255,191}
\definecolor{DeepSnow}{RGB}{191,255,255}
\definecolor{SoapStone}{RGB}{218,218,218}
\definecolor{LightCayenneSixty}{RGB}{239,206,211}
\definecolor{LightCayenne}{RGB}{208,143,145}

\algrenewcommand\algorithmicrequire{\textbf{Input:}}
\algrenewcommand\algorithmicensure{\textbf{Output:}}

\newif\ifcoverletter
\coverletterfalse

\usepackage[most]{tcolorbox}

\definecolor{TakeawayAccent}{RGB}{193,84,20}
\definecolor{ImplicationAccent}{RGB}{20,84,166}

\tcbset{
  calloutbox/.style={
    enhanced,
    breakable,
    sharp corners,
    colback=white,
    colframe=#1,
    boxrule=0.5pt,
    leftrule=2pt,
    boxsep=0pt,
    left=5pt, right=5pt, top=3pt, bottom=3pt,
    before skip=5pt plus 1pt minus 1pt,
    after skip=5pt plus 1pt minus 1pt,
  },
}

\newcounter{takeaway}
\newtcolorbox{takeawaybox}{%
  calloutbox=TakeawayAccent,
  before upper={\refstepcounter{takeaway}%
    \textbf{\textcolor{TakeawayAccent}{Takeaway~\thetakeaway:}}\ },
}

\newcounter{implication}
\newtcolorbox{implicationbox}{%
  calloutbox=ImplicationAccent,
  before upper={\refstepcounter{implication}%
    \textbf{\textcolor{ImplicationAccent}{Implication~\theimplication:}}\ },
}

\begin{document}

%
    \ifcoverletter
        \input{section/cover_letter}%
        \setcounter{page}{1}
    \fi

%
%
\title[]{Dissecting How Die Scaling Breaks GPU \\ Fine-grained Scheduling}

\author{Xiaoze Fan}
\email{jasonfxz@sjtu.edu.cn}
\affiliation{%
  \institution{Shanghai Jiao Tong University}
  \country{}
}

\author{Jianhao Wang}
\email{sword-k@sjtu.edu.cn}
\affiliation{%
  \institution{Shanghai Jiao Tong University}
  \country{}
}

\author{Weihao Cui}
\email{weihao@sjtu.edu.cn}
\affiliation{%
  \institution{Shanghai Jiao Tong University}
  \country{}
}

\author{Han Zhao}
\email{zhaohan\_miven@sjtu.edu.cn}
\affiliation{%
  \institution{Shanghai Jiao Tong University}
  \country{}
}

\author{Zhuobin Huang}
\email{zhuobin@u.nus.edu}
\affiliation{%
  \institution{National University of Singapore}
  \country{}
}

\author{Yangjie Zhou}
\email{yj\_zhou@nus.edu.sg}
\affiliation{%
  \institution{National University of Singapore}
  \country{}
}

\author{Yuxian Qiu}
\email{yuxianq@nvidia.com}
\affiliation{%
  \institution{NVIDIA}
  \country{}
}

\author{Shixuan Sun}
\email{sunshixuan@sjtu.edu.cn}
\affiliation{%
  \institution{Shanghai Jiao Tong University}
  \country{}
}

\author{Bingsheng He}
\email{dcsheb@nus.edu.sg}
\affiliation{%
  \institution{National University of Singapore}
  \country{}
}

\author{Quan Chen}
\email{chen-quan@cs.sjtu.edu.cn}
\affiliation{%
  \institution{Shanghai Jiao Tong University}
  \country{}
}

\author{Minyi Guo}
\email{guo-my@cs.sjtu.edu.cn}
\affiliation{%
  \institution{Shanghai Jiao Tong University}
  \country{}
}

\renewcommand{\shortauthors}{Xiaoze Fan et al.}

\begin{abstract}
Modern GPUs are no longer physically symmetric.
Die scaling leads to both manufacturing-driven floorsweeping and cache and memory partitioning.
The former creates chip-specific compute topologies, while the latter causes non-uniform memory access.
These asymmetries are substantial.
Topology-oblivious compute unit allocation can lead to up to 1.33$\times$ performance variation, while remote accesses increase HBM latency by up to 67\% and nearly double L2 latency.

However, these asymmetries are hidden behind the GPU's logical resource abstractions and can vary across chips.
We develop lightweight characterization methods to uncover per-chip compute topology and memory affinity.
We then use the discovered information to make existing fine-grained scheduling asymmetry-aware, considering not only how many resources are allocated but also which physical resources are assigned.
Across full-GPU kernel execution, intra-application multiplexing, and inter-application co-location, asymmetry-aware scheduling improves mainstream kernels by up to 1.22$\times$, multiplexed LLM inference by up to 14.3\%, and avoids up to 1.33$\times$ performance variation.
\end{abstract}

\maketitle

\section{Introduction}
\label{sec:introduction}

Modern GPUs integrate many parallel resources.
Efficiently utilizing these resources requires fine-grained scheduling at multiple levels.
Kernel libraries schedule work across compute units to approach hardware performance limits~\cite{zadouriFlashAttention4,yeFlashInfer}.
Within an application, recent LLM serving systems spatially multiplex prefill and decode phases to improve goodput~\cite{nvidiaGreenCtx,chenHighgoodput2026}.
Across applications, cloud systems co-locate kernels from different tenants on the same GPU and control each tenant's compute and memory allocation~\cite{nvidiaMPS,wuPTGPUTaskAssignment,zhangSGDRCSoftwareDefined,coppockLithOSOperating,ngPaellaLowlatency,zhaoTally2025}.

Existing fine-grained scheduling works~\cite{zadouriFlashAttention4,yeFlashInfer,nvidiaGreenCtx,chenHighgoodput2026,wuPTGPUTaskAssignment,coppockLithOSOperating,ngPaellaLowlatency,zhaoTally2025} generally operate by controlling logical resource quantities.
For example, they control how many compute units a task uses and how much memory it accesses.
These works omit the physical locations of compute units and memory.
It implicitly assumes architectural symmetry, that allocations with the same number of compute units and the same memory capacity offer equivalent performance.

\begin{figure}
  \centering
  \includegraphics[width=.85\columnwidth]{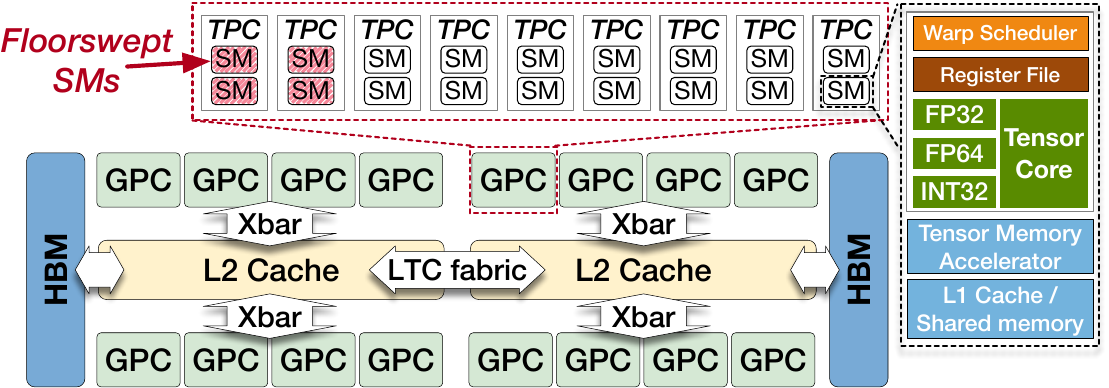}
  \vspace{-1mm}
  \caption{Architecture of an NVIDIA H200 GPU.
  Streaming Multiprocessors (SMs) are grouped into Texture Processing Clusters (TPCs), which in turn form Graphics Processing Clusters (GPCs).
  Each SM has a private L1 cache, while all SMs share a partitioned L2 cache backed by HBM.}
  \vspace{-3mm}
  \label{fig:architecture}
\end{figure}

Die scaling introduces asymmetries, which increases the number of transistors per die by enlarging die area~\cite{nvidiaAmpereArchitecture2020} and shrinking process nodes (7\,nm$\rightarrow$3\,nm)~\cite{tsmc3nmProduction2022}.
\Cref{fig:architecture} shows a modern NVIDIA GPU architecture and how die scaling affects it.
First, floorsweeping disables defective Streaming Multiprocessors (SMs, NVIDIA GPUs' basic compute units) to improve yield, creating chip-specific compute topologies.
Second, larger L2 caches are partitioned for high bandwidth and low latency.
These partitions and their associated HBM form a non-uniform memory access (NUMA) topology with different local and remote access costs.

These asymmetries have substantial performance consequences.
Floorsweeping creates compute asymmetry. Two GPCs can differ by up to 10 SMs on either NVIDIA H200 or B200.
Remote HBM accesses incur approximately 34\% higher latency than local accesses on H200 ($\sim$490$\rightarrow$$\sim$655 cycles) and 67\% on B200 ($\sim$552$\rightarrow$$\sim$920 cycles).
Remote L2 latency is about 51\% higher than local latency on H200 ($\sim$309$\rightarrow$$\sim$466 cycles) and nearly twice as high on B200 ($\sim$364$\rightarrow$$\sim$725 cycles).
As we show later in \S\ref{sec:implications}, ignoring these differences can cause up to 1.33$\times$ performance variation.

This paper addresses two challenges raised by such hidden physical asymmetry.
\textbf{C-1:} How can software efficiently discover hidden and chip-specific compute and memory asymmetries without architectural documentation?
Vendor exposes logical rather than physical SM topology, floorsweeping differs across chips of the same SKU\footnote{SKU (Stock Keeping Unit): a unique product identifier that encodes the GPU product name and interface, e.g., H200.},
NUMA mapping is encoded in undocumented physical-address hashing, and different architectures use different mappings.
Thus, asymmetry often requires per-chip calibration rather than a static lookup table shared across chips of the same SKU.

\textbf{C-2:} How should existing fine-grained scheduling exploit the discovered physical asymmetry?
Existing scheduling works determine only resource quantity.
Asymmetry awareness adds a complementary dimension by considering physical resource identity, topology, and compute-memory affinity.
The goal is not to replace fine-grained scheduling, but to make it more precise by considering not only how many resources are allocated, but which physical resources are allocated.

To address the first challenge, we develop lightweight characterization methods for compute and memory asymmetry.
For compute asymmetry, we uncover the complete per-chip SM-to-GPC assignment using two probing techniques.
We observe two groups of logical SM IDs.
Normal SMs follow an architecture-specific predefined SM-to-GPC mapping, whereas the remaining SMs are assigned to physical GPCs depending on the chip-specific floorsweeping outcome.
For convenience, we refer to the latter as random SMs.
``random'' does not mean runtime-random scheduling, but that their physical GPC locations may vary across chips.
For memory asymmetry, we develop a latency hierarchy discovery method that uncovers the memory partition hash and HBM interleaving granularity on the measured GPUs.
With the uncovered hash, we characterize local/remote latency gaps, remote bandwidth bottlenecks, and effective L2 capacity loss from cross-partition cache replication.

To address the second challenge, we build asymmetry-aware prototypes for three representative fine-grained scheduling scenarios.
For full-GPU kernels, we develop two methods that steers memory accesses to NUMA-local partitions.
Integrating these methods with mainstream GPU kernels improves throughput by up to 1.22\,$\times$ with minimal code changes.
For intra-application multiplexing, we develop a kernel-transparent cross-NUMA allocation method for LLM serving systems that spatially multiplex prefill and decode phases.
This improves decode throughput by up to $14.3\%$ by restricting task-private allocations to NUMA-local partitions while sharing others across partitions.
For inter-application co-location, we evaluate topology-aware SM allocation strategies and show that equal SM counts do not imply equal physical capability.
Topology-oblivious allocation causes up to $1.33 \times$ throughput variation.

We will open-source our code to enable reproduction of all our results.
Our main contributions are as follows:

\begin{itemize}[leftmargin=*, nosep]
    \item We develop lightweight methods to uncover hidden, chip-specific GPU compute and memory asymmetry, including SM-to-GPC mapping, floorsweeping-dependent topology, NUMA partition hashes, and local/remote latency and bandwidth behavior.

    \item We show how fine-grained scheduling can incorporate physical resource identity, topology, and memory affinity rather than relying only on logical resource counts.

    \item We validate asymmetry-aware scheduling across three levels.
    These results motivate making asymmetry awareness a design principle for future GPU programming.

\end{itemize}

\section{Background}
\label{sec:background}

Workloads such as large language models~\cite{grattafioriLlama3Herd2024, jiangMegaScaleScaling2024} demand increasing compute throughput and memory bandwidth, driving aggressive GPU die scaling.
While \Cref{fig:architecture} shows the microarchitecture of a modern GPU die, \Cref{tab:gpu-generations} further summarizes five SKUs spanning four generations of NVIDIA data-center GPUs.
From Volta to Blackwell, full-die SM counts grew from 84 to 160, L2 caches from 6\,MB to 126\,MB, and HBM stacks from 4 to 8.

\begin{table}[t]
  \centering
  \caption{GPU die specifications across generations. H100 ships in SXM
    and PCIe variants; this table lists the PCIe SKU. H100~PCIe and H200
    share the GH100 die but differ in enabled resources. B200 fuses two
    GB100 chiplets; values marked $\times$2 denote per-chiplet quantities,
    while unmarked values are package totals.}
  \label{tab:gpu-generations}
    \vspace{-2mm}
  \scriptsize
  \setlength{\tabcolsep}{2.5pt}
  \begin{tabular}{l |c|c|c|c|c}
    \toprule
    SKU & \textbf{V100} & \textbf{A100} & \textbf{H100} & \textbf{H200} & \textbf{B200} \\
    Die     & GV100~\cite{nvidiaVoltaWhitepaper2017} & GA100~\cite{nvidiaAmpereWhitepaper2020} & \multicolumn{2}{c|}{GH100~\cite{nvidiaHopperWhitepaper2022}} & GB100$\times$2~\cite{nvidiaBlackwellWhitepaper2024} \\
    \midrule
    Die (mm$^{2}$)  & 815  & 826  & \multicolumn{2}{c|}{814} & 800$\times$2 \\
    Full SMs        & 84   & 128  & \multicolumn{2}{c|}{144} & 80$\times$2 \\
    SKU SMs     & 80   & 108  & 114  & 132  & 148 \\
    Disabled SMs    & 4    & 20   & 30   & 12   & 12 \\
    GPCs            & 6    & 7    & 7/8    & 8    & 4$\times$2 \\
    TPCs/GPC        & 7    & 8    & \multicolumn{2}{c|}{9} & 10 \\
    \midrule
    L2 (MB)         & 6    & 40   & 50 & 60  & 126 \\
    L2 partitions   & 1    & 2    & 2 & 2  & 2 \\
    HBM type        & HBM2 & HBM2e & HBM2e & HBM3e & HBM3e \\
    HBM (GB)        & 32   & 80   & 80   & 141  & 90$\times$2 \\
    \bottomrule
  \end{tabular}
  \vspace{-4mm}
\end{table}

\textbf{Compute.}
As shown in \Cref{fig:architecture}, GPU compute is organized hierarchically (GPU - GPC - TPC - SM).
As dies grow, defective SMs must be disabled to maintain manufacturing yield, a process called \emph{floorsweeping}.
The full GH100 die contains 8~GPCs of 9~TPCs each, totaling $8 \times 9 \times 2 = 144$~SMs.
Floorsweeping permanently disables selected TPCs within each GPC, leaving H200 with 132 SMs across 8~GPCs.
Which TPCs are disabled depends on per-die manufacturing defects. After floorsweeping, surviving TPCs are renumbered into a contiguous logical SM~ID space, making the logical-to-physical mapping chip-specific and opaque to software.

\textbf{Memory.}
On the memory side, each SM has a private L1 data cache, and all SMs share a last-level L2 cache backed by off-chip HBM stacks.
SMs communicate with the L2 through a crossbar (Xbar in \Cref{fig:architecture}).
Starting with Ampere, the L2 grew to 40\,MB and was physically split into multiple physical partitions.
On the NVIDIA GPUs studied in this paper, each GPU exposes two memory-affinity partitions.
An LTC fabric connects these partitions to present a unified address space to software.
Newer Blackwell GPUs, such as B200, integrate two chiplets into a single GPU.
Official documentation describes one L2 partition per chiplet, with the two partitions connected by the LTC fabric.
Whether the L2 within each chiplet is further partitioned remains unknown.


\section{Related Work}
\label{sec:related}
\textbf{Fine-grained scheduling of GPU.}
Existing works implement fine-grained scheduling at multiple granularities.
Within a single kernel, persistent thread blocks~\cite{wuPTGPUTaskAssignment,zhao2022ispa,coppockLithOSOperating} allow users to pin thread blocks to specific SMs.
Thread Block Clusters~\cite{nvidiaThreadBlockClusters}, a scheduling feature introduced in Hopper, further leverages this for acceleration.
For multi-task or co-located applications, several systems~\cite{ngPaellaLowlatency,zhaoTally2025,coppockLithOSOperating} schedule multiplexed kernels by controlling the number of SMs.
All these systems control how many SMs are used, but none considers which SMs are assigned or their NUMA affinity.

\textbf{Vendor-supported scheduling.}
NVIDIA's \emph{MPS} (Multi-Process Service)~\cite{nvidiaMPS} supports static SM partitioning.
\emph{Green Contexts}~\cite{nvidiaGreenCtx} provide SM-level partitions with optional GPC alignment.
\emph{MIG} (Multi-Instance GPU)~\cite{nvidiaMIG} provides hardware-isolated compute and memory partitions that can align with NUMA boundaries.
MPS's \emph{MLOPart} (Memory Locality Optimized Partition)~\cite{nvidiaMLOPart} adds NUMA-aware partitioning for Blackwell and newer GPUs.
Both MIG and MLOPart hide detailed physical compute and memory topology.
Moreover, users cannot freely select specific physical SMs or choose between NUMA-local and interleaved placement for each device-memory allocation.

\textbf{Reverse-engineered scheduling.}
FGPUs~\cite{jainFractional2019} and SGDRC~\cite{zhangSGDRCSoftwareDefined} reverse-engineer physical memory mappings and use coloring to reduce interference between co-located workloads.
FGPUs combines page coloring with persistent blocks to reserve SMs and memory bandwidth.
SGDRC dynamically allocates SMs and VRAM channels to improve utilization.
\texttt{libsmctrl}~\cite{bakitaHardwareCompute} enables per-kernel SM partitioning through TPC masking.
The partitioning policies of FGPUs and SGDRC control resource shares and interference without incorporating chip-specific floorsweeping topology or SM-to-memory NUMA affinity into allocation decisions.


\section{Motivation}
\label{subsec:motivation}

NVIDIA publishes the total SM count, L2 cache size, and HBM capacity for each GPU product, but does not disclose the per-chip floorsweeping pattern or the mapping between compute units and memory partitions.
Without this information, software cannot determine which physical resources a logical allocation actually receives.

Understanding this relationship matters at multiple granularities.
At the single-kernel level, thread blocks share SMs and memory partitions, and NUMA-unaware placement can create cross-partition bottlenecks that degrade overall kernel performance.
At the single-application level, developers could overlap internal tasks on the same GPU more effectively if they knew the compute-to-memory affinity.
At the multi-application level, cloud vendors need topology-aware partitioning strategies to ensure fairness across co-located applications.

NVIDIA hides these variable architectural features from software to simplify the runtime and user-level programming model.
For instance, since the introduction of MIG, certain SMs are silently disabled when MIG mode is active, yet no official documentation explains why.
The goal of this paper is to characterize the variable, per-chip architectural asymmetries that die scaling introduces and to show how fine-grained scheduling can use that information.

In the following sections, we first characterize the floorsweeping topology and NUMA affinity mapping.
We then evaluate asymmetry-aware scheduling through several prototype implementations.
We use H200 and B200 as our primary testbeds throughout this paper.

\section{Compute Asymmetry Analysis}
\label{sec:sm-geo}

\subsection{SM Topology Discovery}
\label{subsubsec:sm-topo}
Topology-aware partition placement requires mapping logical SM~IDs to physical GPCs.
We uncover the full mapping with a lightweight, two-phase probing method.

\subsubsection{Phase~1: Cluster probing.}
Thread Block Clusters~\cite{nvidiaThreadBlockClusters} are a Hopper-introduced  new CUDA feature.
They co-schedule a group of thread blocks from a kernel onto \emph{GPC-local} SMs, enabling direct cross-block distributed shared memory access and cluster barriers without going through L2.
In this case, the SMs in the same Thread Block Cluster are guaranteed to be in the same GPC.
To this end, we launch kernels via \texttt{cudaLaunchKernelEx} with cluster sizes ranging from 3 to~8 and log each cluster's constituent SM IDs (obtained via the \texttt{\%smid} PTX register).
SM IDs that co-occur in a cluster belong to the same GPC.
With this method, we can partially uncover the GPC layout of the SMs on the GPU.

For instance, on H200, we can uncover the GPC layout of the SMs with IDs 0--123 (62 TPCs across 8 GPCs).
These SMs reliably appear in cluster launches, while SM~IDs 124--131 (4~TPCs) never appear at cluster sizes~$>$2.
We observe two groups of logical SM IDs.
The SMs identified by cluster probing follow an architecture-specific predefined SM-to-GPC mapping.
For convenience, we call them \emph{normal SMs}.
The remaining SMs have GPC locations determined by the chip-specific floorsweeping outcome.
We call them \emph{random SMs}.
Here, ``random'' does not mean that these SMs are randomly scheduled at runtime.
Rather, their physical GPC locations may vary across chips because of floorsweeping.

\begin{figure}
        \centering
        \includegraphics[width=.9\columnwidth]{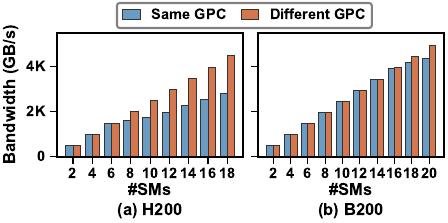}
        \vspace{-3mm}
        \caption{
            Per-GPC L2 bandwidth on H200 and B200 with varied SMs. Because each GPC in Hopper has 9 TPCs, and each GPC in B200 has 10 TPCs, the bandwidth test ends at 18 SMs on H200, and 20 SMs on B200.
        }
          \vspace{-4mm}
\label{fig:h200-b200-gpc-l2-bw}
\end{figure}

\begin{figure*}[t]
    \centering
    \includegraphics[width=1.85\columnwidth]{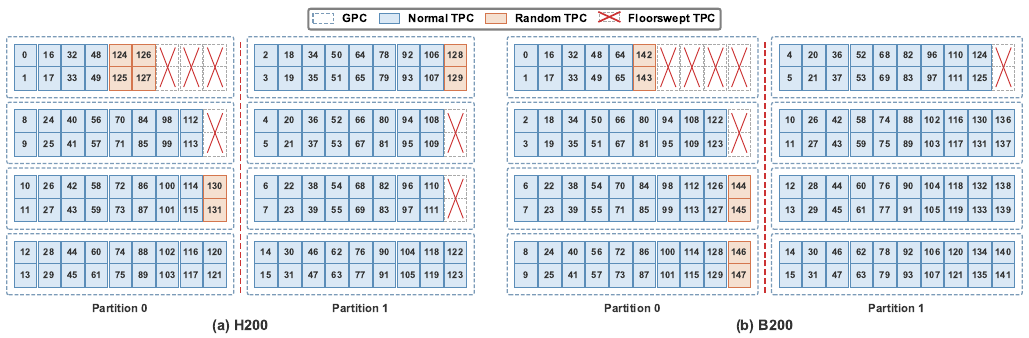}
    \vspace{-4mm}
    \caption{Physical SM layout across GPCs on H200 and B200. Floorsweeping disables different TPCs on each chip. The depicted left--right partition order has no physical significance, and partition~0 can correspond to either side in the actual die shot.}
    \vspace{-3mm}
    \label{fig:gpu-sm-layout}
\end{figure*}

\subsubsection{Phase~2: L2 bandwidth probing.}
\label{subsubsec:l2-bw-probe}
To assign the remaining random SMs, we exploit the observation that SMs sharing a GPC contend on the GPC's L2 Xbar bandwidth~\cite{jinUncoveringReal}.
This means that the SMs in the same GPC have lower L2 bandwidth than the same number of SMs which are distributed to different GPCs.
To get the L2 bandwidth, we use a kernel that repeatedly reads a buffer sized to exceed L1 but fit within L2, forcing all accesses to hit the L2 cache.

\Cref{fig:h200-b200-gpc-l2-bw} shows the L2 bandwidth of SMs within the same GPC and across different GPCs on H200 and B200, respectively.
As shown, the L2 bandwidth grows linearly with the number of SMs when the SMs are located in different GPCs.
However, this trend does not hold for SMs within the same GPC.
When the number of SMs exceeds 6 on H200 and 16 on B200, their L2 bandwidth becomes lower than that of SMs in different GPCs.

With this observation, we can assign the random SMs to the GPCs with the lowest aggregated L2 bandwidth.
Specifically, for each two random SMs from the same TPC with consecutive SM ids, we co-run them with the known SMs of each GPC~$g$ to record aggregate bandwidth.
If adding $s$ causes a bandwidth improvement smaller than a single TPC's bandwidth, $s$ is contending with $g$'s SMs and therefore belongs to~$g$; otherwise $s$ resides elsewhere.
Sweeping all remaining GPCs unambiguously places every random SM.

The combined method runs in under \textbf{1 minute} per GPU and requires only standard GPU kernels.
We validate the uncovered mapping against \texttt{nvdebug} on machines and confirm 100\% agreement across all tested cards.

\begin{takeawaybox}
    Logical SM IDs fall into two groups. Normal SMs follow an architecture-specific predefined SM-to-GPC mapping, while random SMs have GPC locations determined by the chip-specific floorsweeping outcome.
\end{takeawaybox}

\subsection{Chip-specific SM Topology}
\label{subsec:gpc-balance}

\subsubsection{SM Topology Variability}
\Cref{fig:gpu-sm-layout}-(a) shows the fully uncovered SM-to-GPC mapping on a specific H200 and B200 chip, with the blue regions indicating the normal SMs, the orange regions indicating the random SMs, and the part with red crosses indicating the floorswept SMs.
Notably, \Cref{fig:gpu-sm-layout} already shows the affinity of the SMs to the NUMA partitions.
Any 4 GPCs could constitute a NUMA partition.
The logical SM IDs within a NUMA partition are not guaranteed to be the same across different chips.
We will demonstrate the NUMA partition probing in \S\ref{sec:mem-layout}.

By conducting the topology discovery on multiple H200 and B200 chips of the same SKUs, we find that SM counts differ across GPCs within a chip, and GPC configurations vary across chips of the same SKU.
Of the 28 H200 GPUs measured, 23 each have six 16-SM GPCs and two 18-SM GPCs.
Three GPUs each have one 12-SM GPC, three 16-SM GPCs, and four 18-SM GPCs.
One GPU has one 14-SM GPC, four 16-SM GPCs, and three 18-SM GPCs.
The remaining GPU has one 10-SM GPC, two 16-SM GPCs, and five 18-SM GPCs.
Although B200 has two additional SMs per GPC, the GPC imbalance is similar to that on H200.
Additional topology results for the lower-tier H100 PCIe are provided in Appendix~\ref{sec:appendix-h100-pcie}.
Lower-tier products from the same die floorsweep more aggressively and thus expose larger imbalance.
\begin{takeawaybox}
    Due to floorsweeping, SM topology varies within a chip and across chips of the same SKU, producing varied GPC imbalance.
\end{takeawaybox}

\subsubsection{Thread Block Cluster Scheduling}
With the full SM topology uncovered, we re-run the cluster probing experiment on a balanced H200 chip whose 8 GPCs each contain at least 16 SMs.
In this case, one GPC contains 8 normal SMs and 8 random SMs.
Even on this balanced chip, the hardware never schedules blocks with cluster size larger than 2 onto the 8 random SMs.

In principle, these 8 random SMs should support Thread Block Cluster scheduling with cluster sizes larger than 2.
However, the driver or on-device firmware disables random SMs from Thread Block Cluster scheduling.
This hides the floorsweeping induced GPC imbalance from software.
Inspecting CUTLASS~\cite{nvidiaCutlass} confirms this design. Its kernel configuration logic for H200 supports a maximum of 15 thread block clusters with size 8 (14 clusters in 7 intact GPCs + 1 cluster in the smallest GPC).

\begin{takeawaybox}
Thread Block Cluster scheduling is constrained by chip-specific floorsweeping. Random SMs cannot be used for Thread Block Cluster with cluster size larger than 2.
\end{takeawaybox}

\subsection{Scheduling SMs with different mechanisms}
\label{subsec:topo-variability}
With the uncovered SM topology, we can further analyze its impact on the different mechanisms that partition the SMs.

\subsubsection{Priority in Green Context}
We observe Green Contexts prioritizing normal SMs for earlier contexts, leaving random SMs for the last.

Green Contexts partition SMs among kernels running concurrently on the same GPU.
Each kernel is bound to a lightweight Green Context that determines its available SMs.
The driver offers several SM allocation modes for Green Contexts.
By default, the driver allocates SMs in GPC-aligned chunks of 8, tiled across all GPCs so that no allocation concentrates within a few GPCs.
We use the default mode throughout this paper.
The other modes are described in Appendix~\ref{sec:appendix-greencontext}.

Since Thread Block Clusters are widely used in production kernel libraries such as CUTLASS~\cite{nvidiaCutlass}, most deployments require the mode with large cluster size.
This mode introduces an allocation priority across Green Contexts.
Because thread block cluster with size larger than 2 cannot be scheduled onto random SMs, the driver must allocate normal SMs first to maximize the performance for current task.

Therefore, all SMs assigned to earlier allocated Green Contexts are normal SMs, and the last allocated context has the random SMs.
At this time, while the last Green Context may hold the same total SM count as an earlier one, only a subset of its SMs supports Thread Block Clusters larger than 2.
This causes the last context to receive lower performance than expected, because random SMs cannot execute Thread Block Clusters.

\begin{takeawaybox}
To enable large Thread Block Clusters, floorsweeping forces GreenContext to allocate normal SMs first as much as possible, and only allocate random SMs at the end. This creates a priority in SM allocation.
\end{takeawaybox}

\subsubsection{Wasted SMs in MIG}
MIG partitions a GPU into physically isolated instances.
A predefined MIG \emph{profile} specifies the SM count, L2 cache size, and HBM capacity of each instance.
Every chip of the same SKU exposes the same set of profiles.
We find that chip-specific floorsweeping causes MIG to leave some otherwise functional SMs unused.

To examine this loss, we use two H200 MIG profiles, 4g.71\,GB and 3g.71\,GB, abbreviated as 4g and 3g.
They provide 64 and 60 SMs, respectively, each with 30\,MB L2 and 71\,GB HBM.
Their instances can coexist on one GPU, retaining the full memory capacity but leaving 8 of the GPU's 132 SMs unused.

Comparing full-GPU and MIG topologies across chips, we find that each GPC has a fixed SM composition, but the GPCs assigned to each MIG profile vary across chips.
On the H200 in \Cref{fig:gpu-sm-layout}, the 4g instance corresponds to the right-hand partition, containing four relatively intact GPCs.
In the worst case, each of these four GPCs loses two of its 18 SMs to floorsweeping, leaving $4 \times (18 - 2) = 64$ SMs.
The 4g profile must accommodate this worst case across chips.
The depicted chip retains 68 functional SMs in this region, so MIG disables normal SMs 122 and 123 and random SMs 128 and 129 to match the 64-SM profile.

The same reasoning gives $72 - 12 = 60$ SMs for H200's 3g profile if all 12 floorswept SMs fall within its 72-SM region.
It also explains B200's 4g profile, but not its 3g profile.
The 80-SM region for B200's 3g profile could theoretically lose 12 random SMs to floorsweeping, leaving 68 SMs.
However, creating this instance exposes 70 SMs.
One plausible explanation is that manufacturing yield permits a tighter limit of 10 floorswept random SMs in this region, guaranteeing 70 SMs.
Manufacturers control floorsweeping and can impose such limits when defining a SKU.

\begin{takeawaybox}
    Floorsweeping also affects the number of SMs exposed by MIG. To keep profiles identical across chips of the same SKU, MIG sometimes disables even normal SMs.
\end{takeawaybox}

\section{Memory Asymmetry Analysis}
\label{sec:mem-layout}

Starting from Ampere, the L2 cache is physically split into multiple memory-affinity partitions.
This section analyzes the resulting memory layout on modern GPUs.

\subsection{NUMA Layout Discovery}
\label{subsec:mem-topo}

\subsubsection{Access Latency Distribution.}
\label{subsubsec:access-latency-dist}
\begin{figure}
    \centering
    \includegraphics[width=.8\columnwidth]{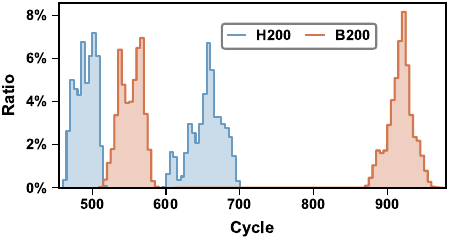}
        \vspace{-3mm}
    \caption{HBM access latency distribution on H200 and B200. H200 exhibits 2 tiers ($\sim$490 and $\sim$655 cycles); B200 exhibits 2 tiers ($\sim$552 and $\sim$920 cycles).}
        \vspace{-4mm}
    \label{fig:h200-b200-2part-latency-hist}
\end{figure}

We first check the latency distribution of the HBM memory access to validate that the split of the L2 cache does introduce the NUMA effect.
With this information, we can identify whether a memory access is local or remote.
To measure the memory access latency distribution, we launch a single-thread kernel on a single SM.
The kernel accesses about 100k randomly sampled addresses across the full address space.
Using this method, we can get the memory access latency distribution of the local and remote HBM memory access.
Kernel implementation details are provided in Appendix~\ref{sec:appendix-hbm-latency-kernel}.

\Cref{fig:h200-b200-2part-latency-hist} shows the HBM access latency distribution on H200 and B200.
We can see there is a clear gap between the local and remote HBM memory access latency.
This validates that the split of the L2 cache does introduce the NUMA effect.
Moreover, both the local and remote HBM access latencies on B200 are higher than those on H200.
In addition, the latency gap between local and remote HBM accesses is also larger on B200.
This could be attributed to B200's larger die size and die-level NUMA structure.

Meanwhile, we also observe that there are twin peaks in the local HBM memory access latency on B200, which stably occurs on all the tested B200 chips.
This suggests that the intra-die L2 cache on B200 is also split into two memory-affinity partitions, which is consistent with the die shot of the B200 chip.
However, the two peaks overlap with each other, making it impossible to distinguish the intra-die partition boundaries from latency alone.
We therefore further analyze whether the intra-die partition matters for bandwidth.

The latency measurement also reveals SM-to-partition affinity, because we could probe from all SMs to each memory partition.
\Cref{fig:gpu-sm-layout} shows the resulting SM layout on an H200 and a B200 chip.
Floorsweeping distributes SM disablement unevenly across the measured NUMA partitions, creating an imbalance in both normal SMs and random SMs.
Across the layouts we observe when creating NUMA-aligned MIG instances, the two H200 partitions differ by up to 12 SMs, and the corresponding B200 partitions differ by up to 8 SMs.

%

\begin{figure}
    \centering
    \includegraphics[width=.8\columnwidth]{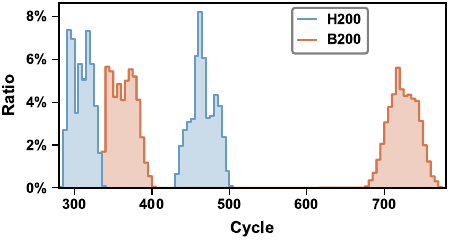}
    \vspace{-4mm}
    \caption{L2 cache access latency distribution on H200 and B200. H200 exhibits 2 tiers ($\sim$309 and $\sim$466 cycles); B200 exhibits 2 tiers ($\sim$364 and $\sim$725 cycles).}
    \vspace{-4mm}
    \label{fig:l2-access-latency-hist}
\end{figure}

Using the SM-to-partition affinity, we measure local and remote L2 cache access latency.
We first cache an address in its local L2 partition, then time accesses from SMs local or remote to that partition while bypassing L1.
Kernel implementation details are provided in Appendix~\ref{sec:appendix-l2-latency-kernel}.

\Cref{fig:l2-access-latency-hist} shows the L2 cache access latency distribution on H200 and B200.
The L2 latency gap shows a similar pattern to the HBM access latency distribution.
B200's local and remote L2 access latencies are both larger than H200's, and the latency gap is also larger.
These observations are consistent with the HBM access latency distribution.

\begin{takeawaybox}
    The split L2 creates a two-node NUMA topology with distinct local and remote latency tiers on both H200 and B200.
    Probing this gap from every SM also reveals SM-to-partition affinity, and floorsweeping leaves the two partitions with unequal SM counts.
\end{takeawaybox}

\subsubsection{NUMA partition identification.}
\label{subsubsec:numa-partition-identification}

\begin{figure}
    \centering
    \includegraphics[width=\columnwidth]{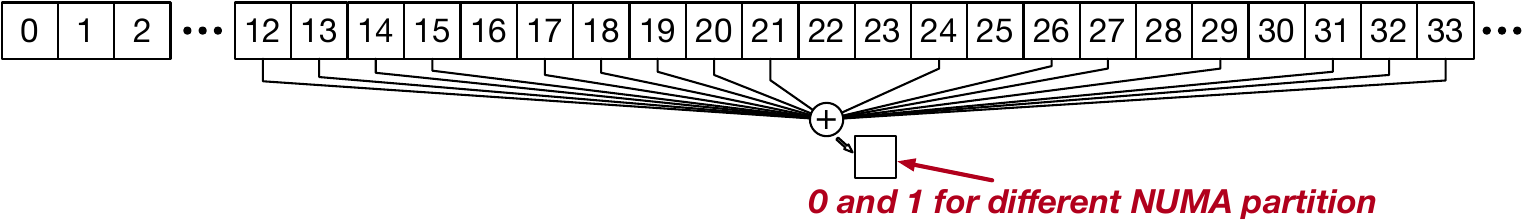}
    \vspace{-6mm}
    \caption{XOR-based hash for NUMA partition on B200.}
    \vspace{-5mm}
    \label{fig:numa-hash}
  \end{figure}

The latency distribution identifies the partition to which a given address belongs.
We also need to determine the partition granularity, defined as the smallest contiguous address range that maps to the same partition.
Knowing the granularity enables us to establish the locality mapping between compute units (SM, TPC, GPC) and memory pages, which is essential for NUMA-aware characterization.

Commonly, a hardware hash function is employed to map each physical address to a partition.
For the two latency-visible partitions, prior work suggests the GPU uses an XOR-based hash over selected physical address bits~\cite{jainFractional2019}.
As shown in \Cref{fig:numa-hash}, the partition hash computes a single parity bit from a subset of physical address bits.
The lowest participating bit determines the granularity, since all addresses differing only below that bit map to the same partition.

We uncover the exact bit positions by flipping each bit individually and observing whether the address switches partitions, measured via latency.
On H200 and B200, 16 bits participate in the hash, with bit~12 as the lowest; on H100~PCIe, 14 bits participate, also starting at bit~12.
Although different SKUs use different partition hashes, the lowest participating bit is consistently bit~12, corresponding to a 4\,KB partition granularity across all three GPUs.

Generally, uncovering the XOR-hash function on a new GPU requires about \textbf{10}\,s, most of which is spent obtaining an accurate HBM access latency distribution.
If verification is required, the time grows linearly with the memory size, taking about 15\,mins for 80GB.

\begin{takeawaybox}
    All dissected GPUs use an XOR hash over physical address bits to map 4\,KB pages across NUMA partitions.
    The hash width and participating bits differ across SKUs, but the 4\,KB partition granularity is consistent.
\end{takeawaybox}

\begin{table*}[t]
    \centering
    \scriptsize
    \setlength{\tabcolsep}{3pt}
    \caption{Three scenarios for asymmetry-aware fine-grained scheduling.}
    \vspace{-3mm}
    \label{tab:asymmetry-aware-scenarios}
    \begin{tabular}{p{0.13\textwidth}p{0.15\textwidth}p{0.17\textwidth}p{0.36\textwidth}p{0.12\textwidth}}
        \toprule
        Usage scenario & Example & Relevant asymmetry & Scheduling action & Key result \\
        \midrule
        Full-GPU kernels & Attention\&MoE  & Memory\&Compute & Topology + NUMA-aware workload assignment & up to 1.22$\times$ \\
        Intra-app multiplexing & Multiplexed prefill/decode & Compute\&Memory & Topology-aware SM + adaptive NUMA allocation & +14.3\% decode \\
        Inter-app co-location & Multi-tenant GEMM & Compute & Topology-aware allocation/order & up to 1.33$\times$ variance \\
        \bottomrule
    \end{tabular}
    \vspace{-2mm}
\end{table*}

\subsection{Performance Impact of NUMA}
\label{subsec:mem-perf}

\begin{figure}
    \centering
    \includegraphics[width=.9\columnwidth]{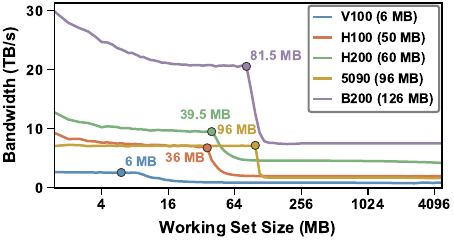}
    \vspace{-3mm}
    \caption{Memory bandwidth as a function of working-set size. The bandwidth cliff marks the effective L2 capacity.}
    \vspace{-5mm}
    \label{fig:all-gpu-bw-vs-blocksize}
\end{figure}
\subsubsection{Effective L2 Cache}
To measure the effective L2 capacity, we adopt an L2 cache benchmark~\cite{gpu-benches} that runs a read-intensive kernel with progressively increasing working-set sizes.
When the working set fits in L2, accesses hit the cache and yield high bandwidth.
Once the working set exceeds the effective L2 capacity, accesses spill to HBM and bandwidth drops sharply.
The inflection point in the bandwidth-vs-working-set curve reveals the effective L2 cache.

\Cref{fig:all-gpu-bw-vs-blocksize} shows the results across five GPU SKUs.
V100 and 5090 have a monolithic L2 cache, while H100, H200, and B200 have a NUMA-structured L2 cache.
For the monolithic GPUs, effective L2 capacity matches the physical L2 size.
For the NUMA-structured GPUs, effective L2 capacity is only 65.8\% of the physical L2 size on H200 and 64.7\% on B200.


With the known NUMA partition granularity, we further measure local and remote memory bandwidth separately.
We classify each 4\,KB page by its NUMA partition within a contiguous memory region, then restrict the benchmark kernel to read only local or remote pages.

\begin{figure}
    \centering
    \includegraphics[width=.9\columnwidth]{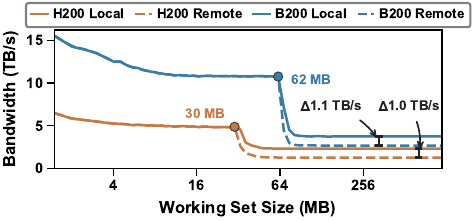}
    \vspace{-3mm}
    \caption{Local vs.\ remote NUMA partition bandwidth on H200 and B200. }
    \vspace{-5mm}
    \label{fig:mig-local-remote-bw}
\end{figure}

\Cref{fig:mig-local-remote-bw} shows the local and remote bandwidth on H200 and B200.
The effective L2 capacity per partition is 30\,MB on H200 and 62\,MB on B200, each half of the respective GPU's total physical L2 capacity.
At the HBM level, a local--remote bandwidth gap persists because the LTC fabric connecting the two L2 partitions limits remote HBM bandwidth.

The effective L2 capacity measurements also reveal B200's intra-die cache behavior.
Although the twin latency peaks in \cref{fig:h200-b200-2part-latency-hist} suggest that each die may contain two L2 partitions, the measured capacity indicates that the L2 within each die behaves as a unified one.
The exploitable NUMA boundary on B200 therefore lies between dies.



\begin{takeawaybox}
    Remote HBM bandwidth is bottlenecked by the LTC fabric that connects the two L2 cache partitions, not by the HBM itself.
\end{takeawaybox}

\subsubsection{L2 cache coherency}
The L2 cache is physically split into memory-affinity partitions connected by the LTC fabric.
This organization raises the question of whether cross-partition access forwards data directly from the remote L2 to the local L1, bypassing the local L2, or replicates data in the local L2 partition using a coherency protocol.

We design a microbenchmark to distinguish these two models.
A local SM first loads a cache line that maps to the remote L2 partition.
A remote SM then invalidates that line.
Finally, the local SM reloads the same address and we measure the access latency.
The reload completes in $\sim$310 cycles on H200 and 360 cycles on B200, matching the local L2 hit latency in \Cref{fig:l2-access-latency-hist}.

This result indicates that the first cross-partition load copied the data into the local L2 partition, where it remained accessible even after the remote copy was invalidated.
Additional load-store experiments confirm this behavior.
Cross-partition L2 access therefore affects local L2 state.
A remote fetch can evict or invalidate existing local L2 entries that alias to the same cache set.

Based on the above analysis, we could also uncover the GPU NUMA hierarchy and access path.
Appendix~\ref{subsec:mem-hierarchy} provides a detailed diagram abouth this.

\begin{takeawaybox}
    The LTC fabric keeps the two L2 partitions coherent by replicating remote data into the local L2 rather than forwarding it to L1.
    Remote data therefore competes for local L2 capacity, which is why NUMA-structured GPUs expose an effective L2 smaller than the physical size.
\end{takeawaybox}

\section{Scheduling Implications }
\label{sec:implications}
Sections~\ref{sec:sm-geo} and~\ref{sec:mem-layout} reveal two architectural asymmetries introduced by GPU die scaling.
However, their scheduling implications depend on how the GPU is used.
We therefore study three representative GPU usage scenarios rather than attempting to build a single monolithic scheduler.

\Cref{tab:asymmetry-aware-scenarios} summarizes these scenarios.
\textbf{In full-GPU kernel execution}, Topology and NUMA-aware memory placement are both evaluated.
\textbf{In intra-application multiplexing}, tasks share the GPU and the scheduler controls both their SM placement and memory placement, making compute and memory asymmetry interact.
\textbf{In inter-application co-location}, independent tenants compete for spatial partitions. In ths case, floorsweeping-induced SM heterogeneity and Thread Block Cluster compatibility directly affect performance isolation and fairness.

\subsection{Full-GPU Kernels: NUMA Locality with Topology Awareness}

We examine how topology- and NUMA-aware workload assignment affects a single kernel occupying the entire GPU.
We first present two NUMA-aware allocation methods that integrate with existing kernels.
Although the kernels occupy all SMs, SM counts per NUMA partition vary across GPUs.
E.g., B200 have three configurations: 70/78, 72/76, and 74/74.
To this end, we first evaluate NUMA awareness alone on GPUs with equal SM counts across partitions, using popular kernels widely deployed in production LLM serving.
We then evaluate combined topology and NUMA awareness on GPUs with unequal SM counts across partitions using a representative kernel.

\subsubsection{NUMA-aware Memory Allocation}
\label{subsub:remapping}

As established in \S\ref{subsubsec:numa-partition-identification}, the GPU interleaves memory across NUMA partitions at 4\,KB granularity.
NUMA-aware execution requires placing data in the same partition as the SMs that process it.
We implement two allocation methods with different architecture coverage and address-computation requirements.

\begin{figure}
    \centering
    \includegraphics[width=.8\columnwidth]{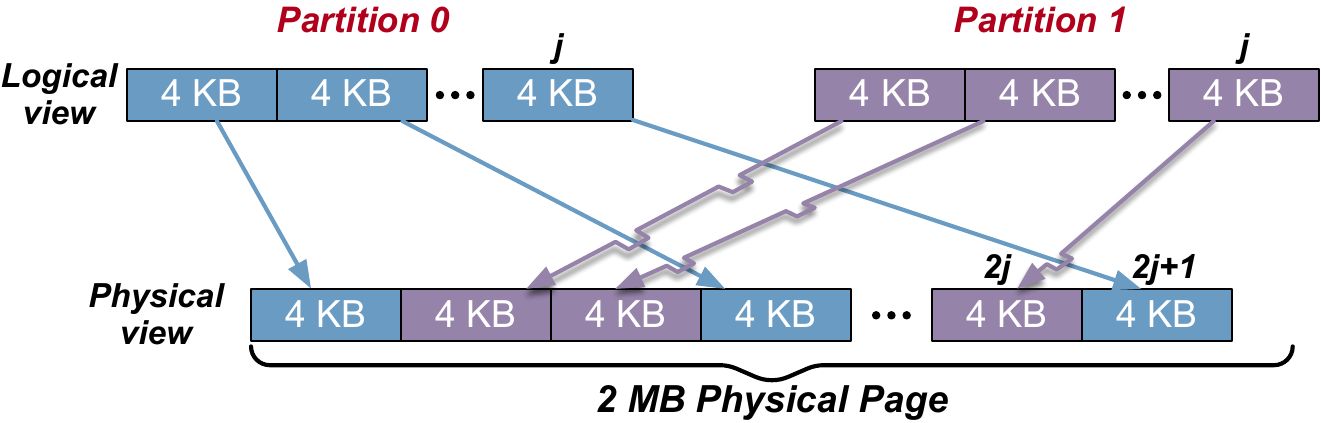}
    \vspace{-3mm}
    \caption{Indirect allocation maps each partition's logical pages to its local 4\,KB pages within a 2\,MB physical page.}
    \label{fig:remap}
    \vspace{-4mm}
\end{figure}

\textbf{Indirect allocation through remapping.}
Our first method supports all NVIDIA GPUs with NUMA.
We first allocate a physically contiguous memory region using large page mappings.
\Cref{fig:remap} illustrates this process within a 2\,MB physical page.
Each aligned pair of consecutive 4\,KB pages contains one page from each partition.
We construct two logical views, each indexing only the pages belonging to its target partition.
To access logical page $j$, the kernel remaps it to one of the two physical pages $(2j,2j+1)$.
The discovered partition hash selects the page belonging to the target partition, while the byte offset within the page remains unchanged.
Both data placement and kernel accesses use this mapping to keep each view's data in its target partition.
This indirection adds an $O(1)$ computation to the kernel's address calculation.
Appendix~\ref{sec:appendix-numa-remapping} provides the implementation details and remapping formula.

\textbf{Direct allocation through driver modification.}
Our second method extends the NVIDIA driver's MLOPart-related code~\cite{nvidiaMLOPart} to expose an allocation interface that accepts a target NUMA partition.
The interface allocates the requested memory within that partition, allowing kernels to access it without software address remapping.
This method removes the remapping overhead but is limited to Blackwell and later GPUs, where the required MLOPart support is available.



\subsubsection{Integration with popular kernels}
\begin{figure}
    \centering
    \includegraphics[width=\columnwidth]{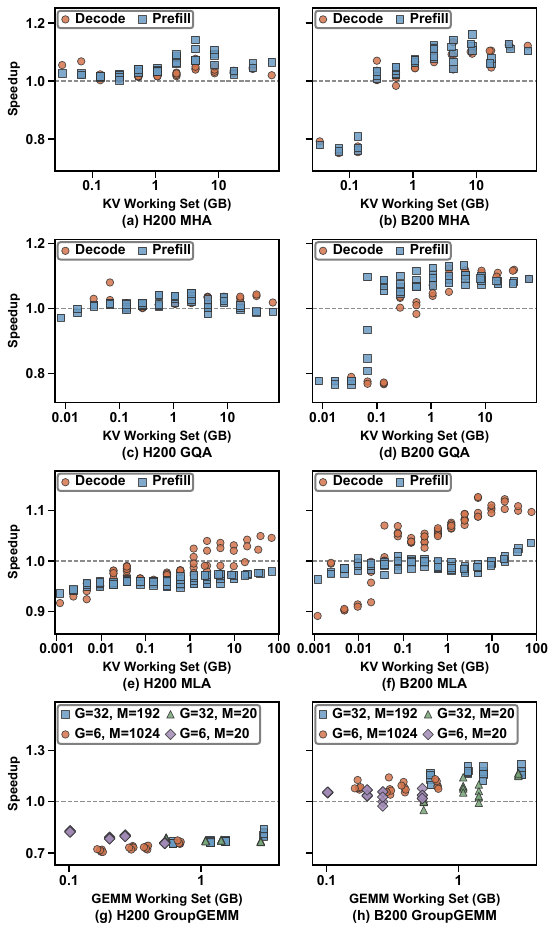}
    \vspace{-4mm}
    \caption{Kernel speedup with NUMA-aware optimizations relative to the corresponding baseline on H200 and B200. For GroupGEMM, G denotes the number of grouped gemms, while M denotes the expected number of rows per group.
    }
    \vspace{-3mm}
    \label{fig:flashattn-single-kernel}
\end{figure}

With either NUMA-aware allocation method, we need to assign the kernel's workload to match its data placement.
We first convert all kernels to use persistent thread blocks (PTB)~\cite{wuPTGPUTaskAssignment}, with one block pinned to each SM to process tasks in a loop.
This fixed block-to-SM mapping allows us to plan in advance which data each block should process.
In this case, we first place the kernel's data in NUMA-local allocations, and then use the discovered SM-to-partition affinity to assign each block the tasks whose data resides in its SM's local partition.
To make the assignment topology-aware, we distribute the data across NUMA partitions in proportion to their SM counts.

We apply NUMA-aware assignment to three attention variants and grouped general matrix multiplication (GroupGEMM), all widely used in LLM.
The attention variants are multi-head attention (MHA)~\cite{vaswaniAttentionAllYouNeed}, grouped-query attention (GQA)~\cite{ainslieGQA}, and multi-head latent attention (MLA)~\cite{deepseekV2}.
GroupGEMM executes multiple independent matrix multiplications within a single kernel.
This structure matches the expert computations in mixture-of-experts (MoE) LLMs, where each expert multiplies its routed token activations by its own weights.
Our baseline implementations come from state-of-the-art kernel libraries.
We use FlashAttention-3~\cite{daoFlashAttention2} on H200 and FlashAttention-4~\cite{zadouriFlashAttention4} on B200 for attention, and CUTLASS on both GPUs for GroupGEMM.
We use indirect allocation on H200 and direct allocation on B200.
We evaluate prefill and decode for all three attention variants with varying key-value (KV) working set sizes, and GroupGEMM with varying GEMM working set sizes.
Benchmark configurations are adopted from the original libraries.
Appendix~\ref{app:full-gpu-config} details the kernel implementations, workload configurations.

\Cref{fig:flashattn-single-kernel} reports speedups over the corresponding baselines on H200 and B200 with equal SM counts across partitions.
Most kernels on both GPUs benefit from NUMA awareness in some configurations, with GroupGEMM on B200 achieving up to 1.22$\times$ speedup.
Hardware counters collected by profilers~\cite{nsightcompute} shows a large reduction in LTC fabric requests across all kernels, confirming that our methods reduce cross-partition accesses.
However, NUMA awareness also degrades performance in some cases.
E.g., attention kernels on B200 with small KV working sets gain little from NUMA awareness.
PTB-based workload assignment~\cite{wuPTGPUTaskAssignment} requires each block to query which SM it is running on, adding startup overhead relative to the original kernel. For small workloads, this overhead outweighs the gains from NUMA awareness, resulting in worse performance.

\begin{figure}
    \centering
    \includegraphics[width=\columnwidth]{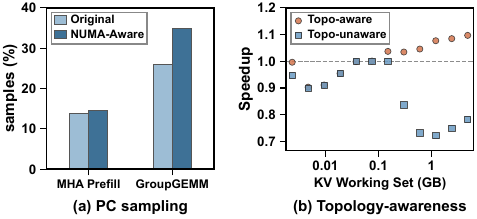}
    \vspace{-6mm}
    \caption{(a) Selected (instruction issue) PC sample fractions for MHA prefill and GroupGEMM on H200, whose increase reflects additional address calculations from NUMA-aware remapping.
    (b) MLA decode speedup with and without compute topology awareness on B200.}
    \vspace{-4mm}
    \label{fig:flashattn-pcsamp-decode}
\end{figure}

Overall, NUMA-aware execution performs better on B200 for two reasons.
B200's cross-die NUMA effect is larger than H200's, and direct allocation incurs negligible address-calculation overhead.
On H200, indirect allocation requires address remapping, which adds overhead, particularly for GroupGEMM.
\Cref{fig:flashattn-pcsamp-decode}-(a) shows program counter (PC) samples associated with address calculation in MHA and GroupGEMM on H200.
This overhead is much larger for GroupGEMM than for MHA, outweighing GroupGEMM's gains from NUMA awareness.

Finally, we evaluate combined topology and NUMA awareness.
\Cref{fig:flashattn-pcsamp-decode}-(b) compares MLA decode performance with and without topology awareness on B200 with unequal SM counts across partitions (70/78).
Without topology awareness, NUMA-aware execution can turn a speedup into a slowdown, from $1.10\times$ to $78\%$.
NUMA locality is important for full-GPU kernel performance, but NUMA-aware workload assignment must also account for SM topology.


\subsection{Intra-Application Multiplexing: Joint Compute--Memory Placement}
\label{subsec:pd-multiplexing}
Intra-application multiplexing co-executes multiple tasks from one application on a GPU, with the scheduler controlling both their SM and memory placement.
A representative example is prefill/decode multiplexing in recent LLM serving systems, which spatially co-execute the two phases to improve throughput~\cite{bulletBoosting,chenHighgoodput2026}.
The phases share weights and KV cache but maintain separate intermediate state, making both compute topology and memory affinity relevant to their placement.

We aim to keep each phase's private state local to its SMs while preserving access to shared weights and KV cache.
However, both allocation methods in \S\ref{subsub:remapping} require kernel changes for asymmetry-aware scheduling, which closed-source libraries in LLM serving stacks prevent.
We instead use two MIG half-instances to run the phases with local and remote memory allocations.
Although MIG does not natively support cross-instance memory sharing, the two phases need access to the same weights and KV cache.
We modify the CUDA driver to allocate physical memory in one instance and map it into the other without changing kernels.
We observed no correctness issues in our tests.

\begin{figure}[t]
    \centering
    \includegraphics[width=.72\columnwidth]{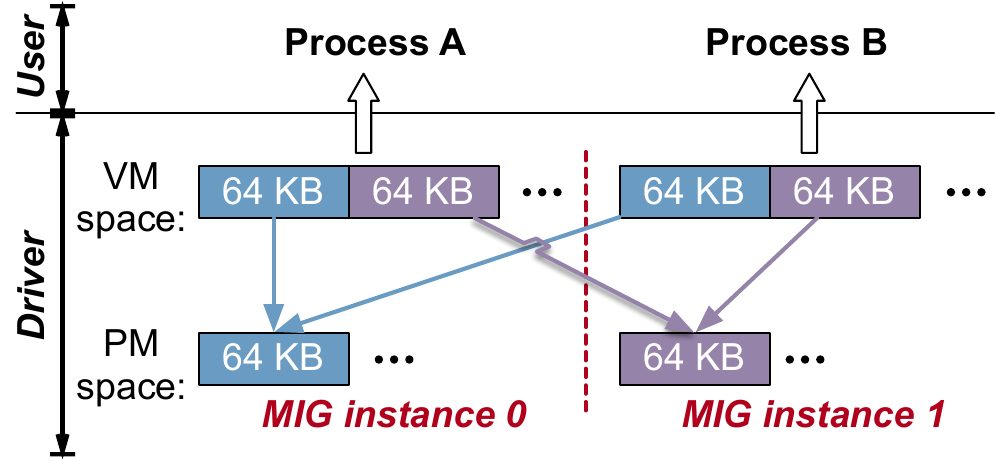}
    \vspace{-2mm}
    \caption{Sharing each MIG instance's local memory between two MIG instances.}
    \vspace{-4mm}
    \label{fig:hack-mig}
\end{figure}

\Cref{fig:hack-mig} illustrates the resulting memory placement.
Each phase's private state resides in its local MIG instance.
Half of the shared memory is allocated locally, and the other half is mapped from the remote instance.
We interleave physical pages across the two instances at 64\,KB granularity.
Our experiments and prior work~\cite{vAttention} confirm that this granularity does not degrade kernel performance.

We integrate this method into mini-SGLang~\cite{miniSGLang} to run Qwen3-8B on H200 and B200.
Each phase can use at most the SMs available in its MIG instance.
Across strategies, prefill and decode receive 64 and 56 SMs on H200, and 64 and 64 on B200, respectively.
These counts are the largest available per MIG half-instance that support kernels with cluster size 8.
On both GPUs, we configure prefill with batch size 4 and sequence length 4K.
Decode uses batch size 128 and average KV length 512.

We compare three multiplexed placements.
\textbf{Cluster-Aware (CA) Overlap} uses Green Contexts with MPS to allocate SMs supporting cluster size 8, without NUMA-aware memory placement, following prior work~\cite{bulletBoosting,chenHighgoodput2026}.
\textbf{Cluster-Mismatched (CM) Overlap} uses the same memory policy, but its SM allocations do not all support cluster size 8.
\textbf{Adaptive-NUMA (AN) Overlap} allocates SMs supporting cluster size 8 and applies the NUMA-aware MIG placement above.
CA and AN also have Standalone configurations, where prefill and decode each run alone with their corresponding placements.
We compare these configurations to verify that our cross-MIG implementation itself does not reduce throughput.

\begin{figure}
    \centering
    \includegraphics[width=.9\columnwidth]{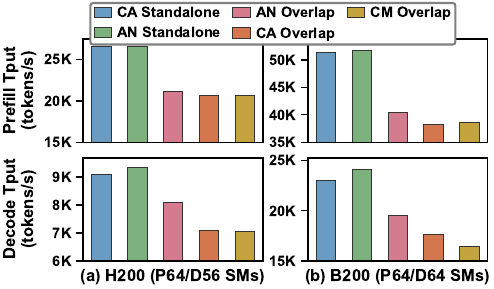}
    \vspace{-2mm}
    \caption{Throughput of prefill and decode under different PD multiplexing strategies on H200 and B200.}
    \vspace{-4mm}
    \label{fig:pdmux-strategies}
\end{figure}

\Cref{fig:pdmux-strategies} presents all the results.
As shown in the figure, CA Standalone and AN Standalone show similar throughput, indicating no evident throughput overhead from the cross-MIG implementation in this configuration.
AN Overlap improves decode throughput over CA Overlap by 14.3\% on H200 and 10.4\% on B200.
These gains come from keeping task-private accesses local, restricting cross-NUMA traffic to shared data.
Prefill gains little because it is compute-bound.

CM Overlap reduces decode throughput by 18.9\% relative to CA Overlap on B200, while the two perform similarly on H200.
Most H200 kernels launch with cluster size 2, so SM topology has limited impact.
On B200, decode kernels perform better with cluster size 8 than with cluster size 2, making cluster-compatible SM placement important.




\subsection{Inter-Application Co-location: Topology-Aware Isolation and Fairness}
In this scenario, we study how the interaction between compute-level asymmetry and Thread Block Cluster affects co-located tenants.
In fine-grained multi-tenancy, compute partitioning can occur at a much finer granularity than physical memory partitioning, so a scheduler cannot always align every tenant's SM allocation with a distinct NUMA partition.
The dominant scheduling problem therefore shifts toward floorsweeping-induced SM heterogeneity, cluster eligibility, and allocation fairness.
Topology-aware compute allocation becomes central.
Equal-sized logical partitions can provide substantially different physical capability.

Thread Block Cluster, introduced in the Hopper architecture, is widely adopted in production libraries such as CUTLASS~\cite{nvidiaCutlass} to accelerate kernels like GEMM.
We extract GEMM kernels from CUTLASS with cluster sizes of 2, 4, and 8, and co-locate each with a baseline GEMM kernel that requires the minimum cluster size of 2.

\begin{figure}[t]
    \centering
    \includegraphics[width=\columnwidth]{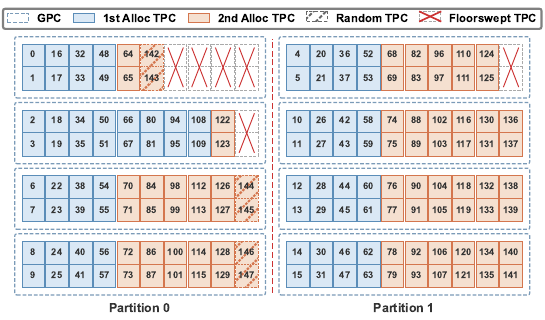}
    \vspace{-5mm}
    \caption{The SM allocation plan created by Green Context on B200.}
    \vspace{-3mm}
    \label{fig:b200-cluster-aware-allocation}
\end{figure}

We denote the two co-located applications as $A$ and $B$. $B$ uses cluster size 2, while $A$ uses cluster size 2, 4, or 8.
On H200, each application requires at least 64 SMs; on B200, at least 72 SMs.
We use Green Contexts with MPS to obtain two SM sets through successive allocations.
In one co-location test, $A$ runs on the first set and $B$ on the second.
In the other, $A$ runs on the second set and $B$ on the first.
\Cref{fig:b200-cluster-aware-allocation} shows an example of the resulting SM allocation on B200.

\begin{figure}
    \centering
    \includegraphics[width=.9\columnwidth]{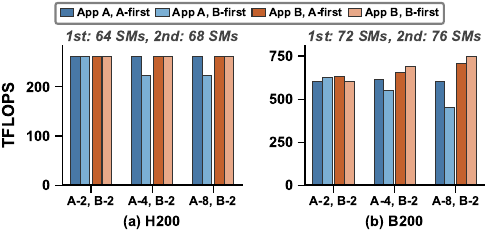}
    \vspace{-2mm}
    \caption{Throughput of different applications launched with varying cluster sizes and SM allocation orders to simulate co-located tenants.}
    \vspace{-4mm}
    \label{fig:cluster-aware-order}
\end{figure}

\Cref{fig:cluster-aware-order} shows the throughput of the two applications in the two cases.
When they use cluster size 2, the two cases yield nearly identical throughput.
For larger cluster sizes, $A$ achieves higher throughput in the case where it runs on the first SM set.
Relative to the other case, the speedups are 1.18$\times$ on H200 and 1.11$\times$ on B200 at cluster size 4.
At cluster size 8, the speedups are 1.18$\times$ on H200 and 1.33$\times$ on B200.

\begin{figure}[t]
    \centering
    \includegraphics[width=\columnwidth]{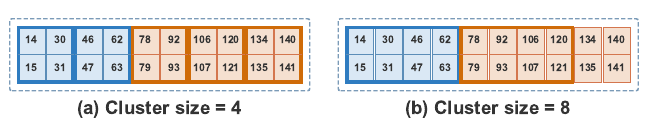}
    \vspace{-4mm}
    \caption{The difference in utilized SMs when launching kernels with cluster sizes 4 and 8 in a B200 GPC with 20 normal SMs.}
    \vspace{-4mm}
    \label{fig:b200-cluster-aware-allocation-one-gpc}
\end{figure}

This gap stems from compute asymmetry.
When $A$ is allocated second, it receives random SMs that cannot form valid clusters, which therefore remain idle during execution.
The effect worsens on B200 at cluster size 8 due to GPC composition.
\Cref{fig:b200-cluster-aware-allocation} shows the detailed SM allocation plan on B200.
The SMs allocated to the first application are marked in blue, while those allocated to the second application are marked in orange.
First, all random SMs are assigned to the second application and cannot be utilized for Thread Block Clusters.
Second, within an intact GPC on B200, the allocation order of $A$ also leads to different outcomes.

As shown in \Cref{fig:b200-cluster-aware-allocation-one-gpc}, each B200 GPC contains 20 SMs.
After $B$ claims 8 SMs from a GPC, 12 SMs remain.
If these 12 SMs are assigned to $A$, they can form three clusters of size 4 but only one cluster of size 8, leaving 4 SMs unused.
This GPC-level asymmetry explains why the throughput penalty increases from 1.18$\times$ to 1.33$\times$ on B200 at cluster size 8.


\subsection{Lessons for Future Fine-Grained Scheduling}

\subsubsection{Full-GPU kernels}
Kernel developers should apply NUMA-local placement when locality gains outweigh PTB and any address-remapping overhead.
Our prototype requires persistent execution and static task assignment using the discovered SM-to-partition affinity.
Developers should distribute data and work in proportion to each partition's SM count to preserve load balance.
In particular, GPU compilers should incorporate NUMA awareness to automate these optimizations across large GPU programs.

\subsubsection{Intra-application multiplexing}
NUMA locality remains important for intra-application multiplexing even when SM allocations are cluster-compatible.
Application runtimes should therefore keep task-private state local and interleave shared state while preserving the cluster compatibility of each phase's SM allocation.
Our cross-MIG prototype implements this policy through driver and runtime changes without modifying kernels.
With current vendor interfaces, our prototype accommodates closed-source kernels through MIG.
However, broader practical adoption will depend on GPU vendors providing mechanisms that enable flexible physical SM selection and per-allocation NUMA placement.

\subsubsection{Inter-application co-location}
Cloud platforms should allow tenants to specify SM topology requirements, such as required Thread Block Cluster sizes, alongside requested SM counts.
Schedulers could then match these requirements to available SM sets when placing co-located workloads to maximize SM utilization.
Such matching would help avoid performance shortfalls caused by SM allocations that cannot support a tenant's topology requirements.
It would also avoid assigning SM sets that support larger clusters to workloads that do not need this capability.

\section{Discussion}

\subsection{Asymmetry in Non-NVIDIA GPUs}
Our characterization methodology also applies to non-NVIDIA GPUs, such as AMD GPUs, which also use die scaling to improve performance.
We investigate AMD GPUs such as MI300X, which have a simpler compute topology than NVIDIA GPUs.
An accelerator complex die (XCD) is analogous to an NVIDIA GPC.
Each XCD contains 40 compute units (CUs), analogous to NVIDIA SMs, with two disabled by floorsweeping.
All XCDs thus have the same number of usable CUs.
These CUs are uniform and do not support hardware features such as Thread Block Cluster.
However, AMD GPUs exhibit NUMA behavior due to their multi-chiplet design.
For example, MI300X integrates four I/O dies within a single GPU, with a memory hierarchy that differs from NVIDIA's.
The partition-local caches are not interconnected by a fabric like NVIDIA's LTC fabric.
AMD GPUs instead integrate an Infinity Fabric linking each XCD to all partition-local caches.
Appendix~\ref{sec:appendix-amd} provides detailed analysis.

Overall, AMD GPUs exhibit memory asymmetry despite their more uniform compute topology.
Thus, we view asymmetry awareness as a key direction for improving fine-grained GPU scheduling performance.

\subsection{Impact on Hardware Simulators}

As in prior hardware architecture characterization work~\cite{huertaDissectingModeling}, our findings could also improve accuracy of GPU simulators by enabling them to model compute and memory asymmetries.
However, such extensions are orthogonal to our focus on using these asymmetries to guide higher-level system design, particularly fine-grained scheduling.
We leave these extensions to future work.

\section{Conclusion}

This paper presents a detailed characterization of the compute and memory asymmetry introduced by GPU die scaling.
Through three prototype case studies, we demonstrate how this asymmetry affects fine-grained scheduling and how awareness of physical topology and memory affinity improves GPU utilization.
We envision that existing GPU programming stacks need to adapt to compute and memory asymmetry to use GPU resources more efficiently.


\bibliographystyle{unsrt}
\bibliography{reference}

\appendix

\section{SM Topology on H100 PCIe}
\label{sec:appendix-h100-pcie}
\begin{figure}[t]
    \centering
    \includegraphics[width=.9\columnwidth]{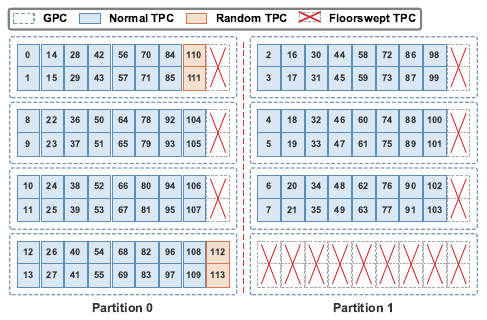}
        \vspace{-3mm}
    \caption{Physical SM layout on an H100 PCIe.}
    \vspace{-4mm}
    \label{fig:h100-pcie-sm-layout}
\end{figure}

To explore the impact of the GPC imbalance on more GPUs, we also conduct the topology discovery on a lower-tier product, H100 PCIe, whose total SMs is 114.
\Cref{fig:h100-pcie-sm-layout} shows the physical SM layout on H100 PCIe.
On H100 PCIe, the SMs numbered from 0 to 109 are the normal SMs, and the SMs numbered from 110 to 113 are the random SMs.
We can see that this H100 PCIe chip has 7 GPCs, with 6 GPCs having at least 16 SMs and 1 GPC having at least 14 SMs, and 1 GPC could be floorswept completely.
In addition, the GPC locations of 4 SMs (2 TPCs) depend on the chip-specific floorsweeping outcome.

\section{Mode in GreenContext}
\label{sec:appendix-greencontext}
In \texttt{IGNORE\_SM\_COSCHEDULING} mode, the driver treats each TPC independently of the GPC hierarchy, enabling fine-grained partitions at the cost of disabling Thread Block Cluster (cluster size $>$ 2).
In \texttt{MAX\_POTENTIAL\_CLUSTER\_SIZE} mode, the driver groups SMs to maximize the achievable cluster size, allocating GPC-aligned chunks of 8 but concentrating them within a few GPCs.

\section{Kernel Implementation Details}
\label{sec:appendix-kernel-implementation}

\subsection{HBM Access Latency}
\label{sec:appendix-hbm-latency-kernel}
To reproduce the HBM latency measurements in \Cref{fig:h200-b200-2part-latency-hist}, we launch a single-thread kernel on one SM to access about 100k randomly sampled addresses across the full address space.
For each address, the kernel first invalidates the L2 line with a PTX instruction \texttt{discard.global.L2} and then issues a PTX \texttt{ld} instruction with the \texttt{.cg} suffix to bypass L1 cache.
Finally, we time the round trip with \texttt{clock64()} to get the global memory access latency.

\subsection{L2 Cache Access Latency}
\label{sec:appendix-l2-latency-kernel}
To measure the local and remote L2 access latencies in~\Cref{fig:l2-access-latency-hist}, we first use an SM $s$ to load an address $p$ mapped to its local partition.
The load uses a PTX \texttt{ld} instruction with the \texttt{.cg} suffix to cache the data in L2 while bypassing L1.
For local latency, we time a second load of $p$ from the same SM $s$.
For remote latency, we instead time the second load from an SM $s'$ with a different NUMA affinity.
The timed loads also use \texttt{.cg} to bypass L1.

\section{Summarized Memory Hierarchy}
\label{subsec:mem-hierarchy}

\begin{figure}
    \centering
    \includegraphics[width=\columnwidth]{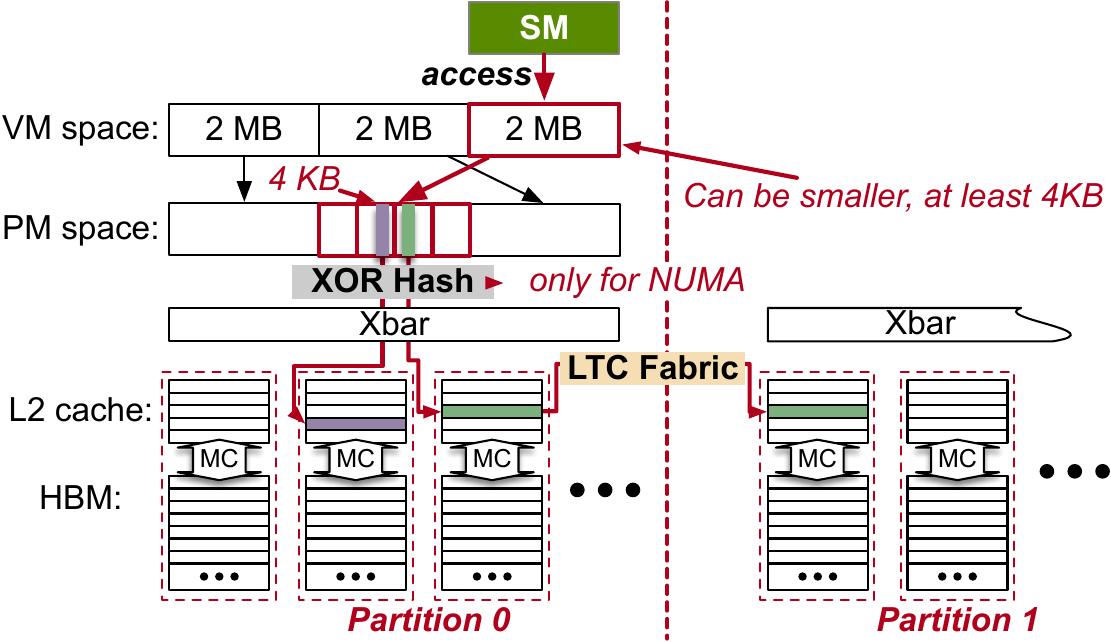}
    \vspace{-2mm}
    \caption{Memory hierarchy layout of a modern GPU. The L2 cache is physically split into memory-affinity partitions backed by dedicated memory controllers and HBM stacks.}
    \vspace{-4mm}
    \label{fig:memory-hierarchy}
\end{figure}

\Cref{fig:memory-hierarchy} summarizes the memory hierarchy of a modern GPU based on our analysis in \S\ref{sec:mem-layout} and prior work~\cite{jinUncoveringReal,zhangSGDRCSoftwareDefined,jainFractional2019}.
When an application allocates memory, the GPU driver maps virtual addresses to physical addresses through page tables at a configurable page size.
A hardware XOR hash maps physical addresses to NUMA partitions at 4\,KB granularity, interleaving memory across partitions to mitigate bandwidth imbalance.
The partition granularity is independent of the page size used for address translation.

When an SM issues a memory request, the GPU translates the virtual address and uses the resulting physical address to determine the target NUMA partition.
For address spaces whose sizes are not powers of two, prior work identifies a separate, non-XOR hash for L2 slice and HBM channel selection~\cite{zhangSGDRCSoftwareDefined}.
Because the NUMA effects studied here occur at the partition level, we omit these lower-level mappings.
If the target page resides in the local partition, the request proceeds through the local L2 slices to the local HBM channels.
For a remote page, the request traverses the LTC fabric to the remote L2 slices and HBM channels.
The fetched data is then copied into the local L2 slices.

\section{NUMA-aware Address Remapping}
\label{sec:appendix-numa-remapping}

The address-remapping method in \S\ref{subsub:remapping} provides partition-local memory access while retaining large page mappings to limit translation lookaside buffer (TLB) pressure.
The 4\,KB NUMA interleaving granularity is independent of the page size used for address translation.
Our initial attempt modified the GPU driver to use a 4\,KB page size for NUMA-aware allocation\footnote{NVIDIA GPUs natively support at least three page sizes: 4\,KB, 64\,KB, and 2\,MB~\cite{nvidiaPascalMMU}.
The default page size is 2\,MB to reduce TLB pressure.}.
The resulting TLB pressure degraded the performance of memory-bound kernels.
We instead retain large page mappings and remap addresses within the kernel at 4\,KB granularity.

Following vAttention~\cite{vAttention}, we modify the NVIDIA driver to allocate physically contiguous memory and expose its base physical address.
The partition hash is an XOR of physical address bits (\S\ref{subsubsec:numa-partition-identification}).
On H200 and B200, the hash mask includes bit~12, the lowest bit of the 4\,KB physical page number.
For an allocation starting at an even physical page number, the two pages in each pair $(2j,2j+1)$ differ only in bit~12.
As shown in~\Cref{fig:remap}, these pages therefore hash to different partitions, so each pair contains exactly one page from each partition.

For an $S$-byte allocation comprising complete page pairs, each partition holds $S/2$ bytes.
To map logical page $j$ to the page in pair $(2j,2j+1)$ belonging to target partition $P$, we compute:
\begin{equation}
\textit{page}_{\textit{phys}} = 2j + \bigl[\text{parity}\bigl((\textit{page}_{\textit{start}} + 2j)\;\&\;\textit{MASK}\bigr) \oplus P\bigr]
\label{eq:remap}
\end{equation}
\noindent where $\textit{page}_{\textit{start}}$ is the allocation's starting physical page number, $\textit{MASK}$ is the GPU-specific partition bitmask shifted to page granularity, and $P \in \{0,1\}$.
The result $\textit{page}_{\textit{phys}}$ is a page offset relative to the allocation base, and the byte offset within the page is unchanged.
The partition selection uses bitwise AND, popcount parity, and XOR, providing $O(1)$ address translation without a lookup table.
This mapping lets the SMs in each NUMA partition process the corresponding $S/2$ portion of the total $S$-byte workload using local pages.

\section{Full-GPU Kernel Experimental Configurations}
\label{app:full-gpu-config}

\paragraph{Hardware}
\Cref{fig:flashattn-single-kernel} uses NVIDIA H200 and B200 devices with 132 and 148 SMs, respectively. The two NUMA partitions contain 66/66 SMs on H200 and 74/74 SMs on B200. The topology-awareness experiment in \Cref{fig:flashattn-pcsamp-decode}-(b) uses a B200 device with 70/78 SMs in its two partitions.

\paragraph{Workloads}
For the attention workloads, we denote the query and KV sequence lengths by $S_q$ and $S_k$, respectively, measured in tokens per request.
Prefill processes $S_q=128$ query tokens per request against a KV sequence of length $S_k$, while decode processes $S_q=1$ query token per request.
Across the attention experiments, batch sizes $B$ range from 1 to 64 for both phases.

For each batch size, we sweep $S_k$ from 512 tokens by doubling.
The attention plots report the logical KV working set in decimal GB:
\begin{align*}
 W_{\rm MHA/GQA} &= 4 B S_k H_{kv}D/10^9,\\
 W_{\rm MLA} &= 2 B S_k(512+64)/10^9,
\end{align*}
where $H_{kv}$ is the number of KV heads (32 for MHA and 8 for GQA), and $D=128$ is the head dimension for both.
MLA counts one compressed KV representation without an additional head-count or K/V duplication factor.

For GroupGEMM, we test $(G,M)=(32,192)$, $(6,1024)$, $(32,20)$, $(6,20)$.
Each setting uses four matrix shapes: $(N,K)=(6144,7168)$, $(7168,3072)$, $(4096,4096)$, $(4096,2048)$, for 16 configurations in total.
For each configuration, we independently sample the actual row count of each group as $M_i=\lfloor M U_i\rfloor$, where $U_i\sim\mathrm{Uniform}(0.7,1.3)$ for $i=1,\ldots,G$.
Five shape seeds per configuration yield 80 instances per GPU, plotted individually.
The GroupGEMM plots report the logical working set of the input and output matrices in decimal GB:
\[
 W_{\rm GEMM} = 2\bigl[(\textstyle\sum_i M_i)(K+N)+GNK\bigr]/10^9.
\]

For the H200 PC-sampling comparison in \Cref{fig:flashattn-pcsamp-decode}-(a), we use MHA prefill with $B=64$, $S_q=128$, and $S_k=65{,}536$, and GroupGEMM with $G=32$, $M=192$, and $N=K=4096$.

\paragraph{Profiling setup.}
We collect hardware counters and PC samples using NVIDIA Nsight Compute (NCU)~\cite{nsightcompute}.
To examine the address-remapping overhead on H200 discussed in the main text,
we use PC sampling to compare the original and NUMA-aware kernels in \Cref{fig:flashattn-pcsamp-decode}-(a).
The figure reports the fraction of samples in the Selected state, which indicates that a warp issued an instruction.
Address remapping adds arithmetic and bitwise operations, which can increase the fraction of samples in which warps issue instructions rather than wait for memory.
The Selected share increases from $25.96\%$ to $35.04\%$ for GroupGEMM, compared with $13.92\%$ to $14.71\%$ for MHA, supporting the conclusion that address remapping adds more overhead to GroupGEMM than to MHA.

\section{Results on AMD GPUs}
\label{sec:appendix-amd}

We characterize memory access asymmetry on an AMD Instinct MI300X GPU.
We first describe its memory organization, then examine how access
latency varies with the requesting XCD and the target address.

\begin{figure}[t]
    \centering
    \includegraphics[width=\columnwidth]{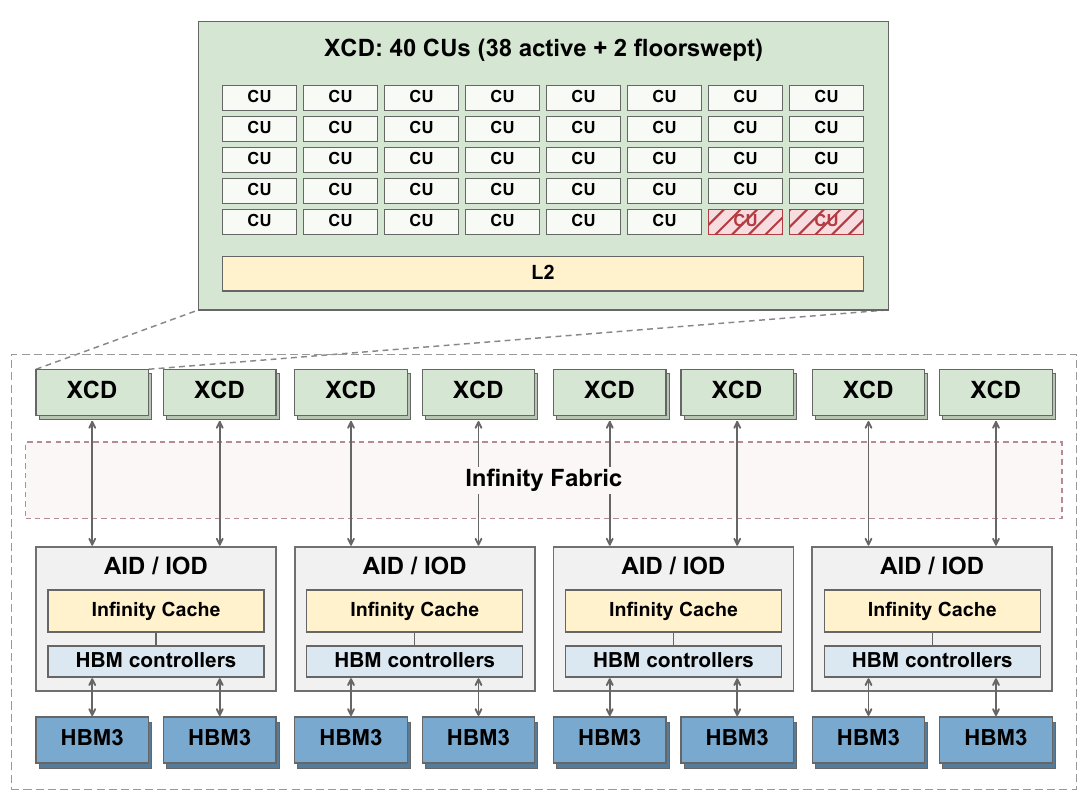}
    \caption{Simplified MI300X architecture. Each AID supports two stacked
    XCDs and connects to two HBM3 stacks. The vertical layout illustrates
    component relationships: HBM stacks physically sit beside the AIDs,
    and Infinity Fabric denotes connectivity rather than a separate
    physical layer.}
    \label{fig:mi300x-architecture}
\end{figure}

\paragraph{Memory organization.}
Figure~\ref{fig:mi300x-architecture} illustrates the MI300X package.
It contains eight XCDs and four active interposer dies (AIDs), also
called I/O dies (IODs). Each AID supports two vertically stacked XCDs
and connects to two HBM3 stacks~\cite{amd_smi_partitioning}.
Each XCD contains a shared 4\,MB L2 cache. The AIDs contain the HBM controllers and slices of Infinity Cache, a memory-side cache with 64\,MB per AID and 256\,MB in total.
Infinity Cache resides on the AIDs, rather than in the HBM stacks.
Infinity Fabric provides the interconnect linking the compute and I/O dies: accesses leaving an XCD's L2 traverse the fabric to reach the memory-side resources
on the IODs~\cite{amd_cdna3_whitepaper}. This organization motivates examining whether different compute--memory locations exhibit different access latencies.

\paragraph{Access latency distribution.}
Following the latency-based characterization in \S\ref{subsubsec:access-latency-dist}, we measure the HBM access latency distribution to examine NUMA effects on MI300X.
We adapt the single-thread latency probe in Appendix~\ref{sec:appendix-hbm-latency-kernel} to AMD, bypassing L1/L2 and flushing Infinity Cache between passes.
We probe the same one million randomly sampled, 128-byte-aligned addresses within a 32\,GiB allocation from each XCD. For each address and XCD, we retain the minimum latency over eight passes.

\begin{figure}[t]
    \centering
    \includegraphics[width=\columnwidth]{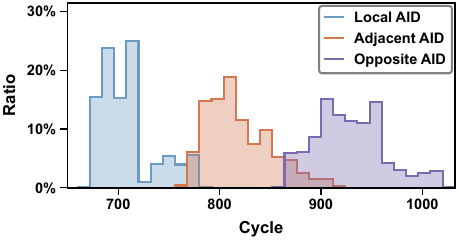}
    \caption{HBM access latency distribution on MI300X from XCD~0.
    Local, Adjacent, and Opposite accesses have mean latencies of
    approximately 707, 817, and 926 cycles, respectively. Each series
    is normalized independently using all of its samples.}
    \label{fig:mi300x-latency}
\end{figure}

Figure~\ref{fig:mi300x-latency} shows the HBM access latency distribution from XCD~0, with addresses classified by their memory affinity as described below. We observe three distinct latency tiers, with mean latencies of approximately 707, 817, and 926 cycles.
The gap between these tiers shows that memory access latency depends on the affinity between the requesting XCD and the target address. H200 and B200 exhibit two main local and remote latency tiers in \S\ref{subsubsec:access-latency-dist}. On MI300X, remote accesses further separate into two tiers, approximately 110 and 219 cycles above the local mean.

\paragraph{XCD-to-memory affinity.}
As the SM-level probes in \S\ref{subsubsec:access-latency-dist} reveal SM-to-partition affinity,
probing the same addresses from different XCDs reveals XCD-to-memory
affinity on MI300X. We group XCDs by correlations in their
address-dependent latency patterns and identify four pairs:
$\{0,1\}$, $\{2,3\}$, $\{4,5\}$, and $\{6,7\}$.
For each address, we infer its home group as the pair with the lowest mean access latency. Each group accounts for approximately 25\% of the sampled addresses. This four-pair structure is consistent with the two-XCD-per-AID organization in Figure~\ref{fig:mi300x-architecture}.

From XCD~0, addresses assigned to its own group form the Local series.
The two remote groups with intermediate mean latency form the Adjacent series, while the group with the highest mean latency forms the Opposite series.
These names denote inferred locality classes rather than measured fabric hop counts.

The measurements therefore reveal four memory-affinity groups and three access-latency tiers on MI300X. Together with the H200 and B200 results in \S\ref{subsec:mem-topo}, they show that latency-based probing can reveal compute--memory affinity across different GPU memory organizations.

\end{document}